# Crystal structure prediction via combining graph network and Bayesian optimization

Guanjian Cheng[1], Xin-Gao Gong[2,3], Wan-Jian Yin[1,4,5]

[1]College of Energy, Soochow Institute for Energy and Materials InnovationS (SIEMIS), and Jiangsu Provincial Key Laboratory for Advanced Carbon Materials and Wearable Energy Technologies, Soochow University, Suzhou 215006, China
[2]Key Laboratory for Computational Physical Sciences (MOE), State Key Laboratory of Surface Physics, Department of Physics, Fudan University, Shanghai 200433, China
[3]Collaborative Innovation Center of Advanced Microstructures, Nanjing 210093, China
[4]Light Industry Insititute of Electrochemical Power Sources, Soochow University, Suzhou 215006, China
[5]Key Lab of Advanced Optical Manufacturing Technologies of Jiangsu Province & Key Lab of Modern Optical Technologies of Education Ministry of China, Soochow University, Suzhou 215006, China

**Abstract**
We developed a density functional theory-free approach for crystal structure prediction via combing graph network (GN) and Bayesian optimization (BO). GN is adopted to establish the correlation model between crystal structure and formation enthalpies. BO is to accelerate searching crystal structure with optimal formation enthalpy. The approach of combining GN and BO for crystal Structure Searching (GN-BOSS), in principle, can predict crystal structure at given chemical compositions without additional constraints on cell shapes and lattice symmetries. The applicability and efficiency of GN-BOSS approach is then verified via solving the classical Ph-vV challenge. It can correctly predict the crystal structures of 24 binary compounds from scratch with averaged computational cost ~ 30 minutes each by only one CPU core. GN-BOSS approach may open a new avenue to data-driven crystal structural prediction without using the expensive DFT calculations.

Predicting crystal structure at given chemical composition *in prior* to experimental synthesis has long-lasting interest in condensed matter science. Earlier attempts based on empirical rules give qualitative description of structures, such as Pauling's five rules for ionic crystals [1], Goldschmidt's tolerance factor for perovskite formability [2], and dimensional descriptors to classify the zinc-blend(ZB)/wurtzite(WZ) and rock-salt (RS) structures for binary semiconductor compounds [3,4]. Owing to reliable energy calculation via density functional theory (DFT), the current state-of-the-art approaches for crystal structure prediction (CSP) are mainly combining DFT calculations with structural searching algorithms such as (quasi-)random search [5,6], simulating annealing [7,8], genetic algorithm [9,10], particle-swarm optimization (PSO) [11] and differential evolutionary process [12]. Those approaches extensively explore the structural candidates via searching algorithms, adopting DFT-calculated energy as stability metric. The necessary DFT calculations are the most time-consuming part as evaluating numerous structural candidates in the process of structure searching.

The renaissance of machine learning (ML) in materials science by the advent of materials database intrigues much efforts on its applications in predicting materials properties such as formation enthalpies ($\Delta H$) [13], Gibbs free energies [14], bandgaps [15], and even wave function and electron density [16], X-ray absorption spectra [17], phase transitions [18] with accuracy close to quantun mechanic calculations yet computational cost orders of magnitude faster. In additional to compositional atoms, the influence of their spatial arrangement, *i.e.,* crystal structure, on materials properties have recently been considered via structural characterizing approaches such as Wyckoff-species matrix-based method [19], Voronoi tessellation method [20], and graph network [21,22]. A crystal **G** can be represented by a vector $\left(\{\mathbf{C}_i\}_{i=1,N},\{\mathbf{R}_i\}_{i=1,N},\mathbf{L}\right)$, where $\{\mathbf{C}_i\}$ and $\{\mathbf{R}_i\}$ are elemental features and coordinates of the *i*th atom, $N$ is the total number of atoms in a periodic cell and **L** is the vector defining cell shape. In those approaches, crystal structures are transformed to physically meaningful and algorithm readable data format, such as symmetry-invariant matrix [19], bond configurations [20] or crystal graphs [21], enabling the establishment of correlation model between crystal and its formation enthalpy,

$$\Delta H = f\left(\mathbf{G}(\{\mathbf{C}_i\},\{\mathbf{R}_i\},\mathbf{L})\right) \qquad (1)$$

In principles, CSP can be efficiently performed according to Eq. (1) by optimizing $(\{\mathbf{R}_i\},\mathbf{L})$ to minimize $\Delta H$ at given $\{\mathbf{C}_i\}$. This approach substitutes DFT calculations with ML model, therefore, has potential to significantly accelerate the CSP.

Despite the promise, the practical approach of DFT-free CSP still have challenges: *(i)* the ML model should have sensitive response to crystal structure, therefore, the fixed-structure model [13] or symmetry-invariant model [19] with constraint on crystal structures are inapplicable or limited in determining the ground state crystal structure that may have arbitrary cell shape and atomic coordinates; *(ii)* the high accuracy of DFT calculations benefit from a systematic cancellation of errors relative to experiment and the claimed DFT-level accuracies of ML models were obtained on training data composed of stable crystal structures [23]. The extension of ML models to structural searching are questionable, since most structural candidates in searching process are metastable or unstable and their relative energies are crucial ingredient to determine the ground state structure; *(iii)* An appropriate optimization algorithm compatible to *black-box* ML model is required.

In this Letter, we construct an improved graph network to establish the ML model between ~320,000 crystal compounds and their DFT-calculated formation enthalpies, with MAE as less as 26.9 meV/atom. This model is then combined with Bayesian optimization (BO) for CSP. The combined approach of Graph Network-based Bayesian Optimization Structural Searching (GN-BOSS) was then applied to successfully predict the crystal structures of 24 octet binary compounds, *i.e.*, I-VII (I=Li, Na, K, Rb, Cs; VII=F, Cl) and II-VI (II=Be, Mg, Ca, Sr, Ba, Zn, Cd; VI=O, S). The comparative studies show that DFT-free GN-BOSS approach is able to predict crystal structures from scratch with extremely low computational cost. This work may open a new avenue to data-driven crystal structural prediction without using the expensive DFT calculations during structural searching.

A GN model has been firstly constructed to establish the correlation between crystals and their formation enthalpies $\Delta H$. In GN, crystal structures are considered as a crystal graph **G**, numerically represented by atomic and structural features [21,22]. We initially used twenty-three atomic features {$C_i$} and ten structural features. The structural features contain six cell features {$a, b, c, \alpha, \beta, \gamma$} (abbreviated as **L**) and four kinds of attributes of Voronoi polyhedrons [24,25] derived from atomic coordinates {$R_i$}. The advantage of using GN to describe crystal is that it has no constraints on cell shape and lattice symmetry and there are well-developed update operations, concatenation and aggregation for **G** [26], which help map an arbitrary crystal graph **G** to $\Delta H$ [Eq. (1)]. Since the DFT-calculated $\Delta H$ depends on the setup parameters, we also considered three key setup parameters as features during GN model training. The details of features (Table S1-S3) and of GN architecture are shown in Supplementary Materials.

GN was updated based on DFT-calculated $\Delta H$ from the offline Open Quantum Materials Database (OQMD) [27]. Data cleaning was performed to exclude data with incomplete information and with restrictions: *i*) number of atoms in the unit cell (less than 50); *ii*) PBE as exchange-correlation functional; *iii*) kinetic energy cutoff (520 eV), making data as reliable and comparable as possible. Accordingly, we obtained more than 320,000 data points, including ~40,000 experimental known ones and ~280,000 hypothetical ones, covering 85 elements, 7 lattice systems and 167 space groups. All data was divided into three parts - training set (50%), validation set (12.5%) and test set (37.5%). The best model was chosen to minimize the errors between GN-predicted and DFT-calculated Δ*H*. Previous studies [21,22] show that the network architecture and update functions have large impact on the model performance. Here, we incorporated the MEGNet architecture [22] (Figure S1) by the compositional features [28] and structural features from Voronoi polyhedrons [29]. MEGNet architecture was adopted since it is a universal framework for crystals and molecules [22], facilitating future extension of current approach to molecular structure prediction. Notably, our GN models can reach MAE of 26.9 meV/atom when the portion of training data is 50% (Figure 1), less than previous model of CGCNN (39 meV, MP database), iCGCNN (30.5 meV, OQMD database) and original MEGNet (28 meV, MP database). We find a dramatic decrease of MAE as the number of training data increase and the values is rapidly converged below 50 meV when the portion of training data reaches as less as 25%, showing the generality of current GN model.

In the framework of GN, CSP becomes an optimization problem to identify the crystal graph $\mathbf{G}(\{\mathbf{R}_i\},\mathbf{L})$ at given chemical composition $\{\mathbf{C}_i\}$ to minimize Δ*H* according to GN-constructed Eq. (1). Enumeration of all possible structures is a long-standing challenge in CSP. Here, we apply BO via a Tree of Parzen Estimators (TPE) -based Gaussian mixture model [30] to explore the structural space. BO is a selection-type algorithm that is compatible to black-box ML model, demonstrating great capability to identify the global minimum [31] and was recently combined with DFT calculations for CSP in fixed crystal systems [32]. In comparison to normal BO algorithm based on Gaussian process which performs better in low-dimensional space (number of features less than 20), TPE-based Gaussian mixture model demonstrated higher efficiency in high-dimensional space [30]. Here, we combined GN model with BO algorithm as GN-BOSS approach, as shown in Figure 2, to do minimization problem, *i.e.*, $\min_{\{\mathbf{c}_i\}} f\left(\mathbf{G}(\{\mathbf{C}_i\},\{\mathbf{R}_i\},\mathbf{l})\right)$.

Firstly, *n* initial random structures $\mathbf{G}_k(\{\mathbf{R}_i\},\mathbf{L})_{k=1,n}$ at given $\{\mathbf{C}_i\}$ were generated and their corresponding structural features were obtained by Voronoi analysis [20]. Accordingly, Δ*H*'s are predicted as $f\left(\mathbf{G}_k(\{\mathbf{R}_i\},\mathbf{L})\right)$ by our GN model. Then, BO algorithm would build surrogate functions based on vectors $\left[f\left(\mathbf{G}_k(\{\mathbf{R}_i\},\mathbf{L})\right)\right]_{k=1,n}$. TPE-based Gaussian mixture model and acquisition functions have been used to suggest possible global minimum at $(\{\mathbf{R}_i\}, \mathbf{L})_{n+1}$. Then, its *'true'* formation energy $f\left((\{\mathbf{R}_i\}, \mathbf{L})_{n+1}\right)$ is then obtained by our GN model and iteratively feedback to refine the BO model to suggest (*n*+2)*th* structure $\mathbf{G}(\{\mathbf{R}_i\}, \mathbf{L})_{n+2}$. For practical implementation, we added additional constraint to avoid generation of unreasonable structures with extremely small or large volume. Notably, surrogate functions in BO model is a kind of approximation to the black-box function for Eq. (1), based on limited (**G**, Δ*H*). With additional **G** in model training as shown in Figure 2, the BO model is expected to become more and more accurate.

The GN-BOSS approach was then applied to solving typical Ph-vV challenge [3,4,33], *i.e.* identify the crystal structures of octet binary compounds, including 10 I-VII (I = Li, Na, K, Rb, Cs; VII = F, Cl) compounds and 14 II-VI compounds (II = Be, Mg, Ca, Sr, Ba, Zn, Cd; VI = O, S). There are more than 300 kinds of prototype structures for AB compounds [27], two representatives of which are tetrahedral-coordination ZB/WZ and octahedral-coordination RS structures. Discerning ZB/WZ and RS structures proves the ability of CSP from ionic to covalent systems [4]. The GN model is trained based on the data excluding 24 compounds above. Here, we are taking four-atom cell CaS for example to show how GN-BOSS works. Firstly, 200 structures with random $\mathbf{G}(\{\mathbf{R}_i\}, \mathbf{L})_{i=1,4}$ have been generated and then BO was iteratively performed starting from that to choose low-Δ*H* structure, with their GN-predicted Δ*H* and typical structures shown in Figure 3. To show the performance of BO, random searching (RAS) have also been shown in Figure 3 for comparison. To verify the extension of GN model to the structural candidates in searching procedure, we take out structures every 50 RAS steps and use DFT to calculate their total energies. The DFT-calculated total energies follow the similar trend as GN predicted Δ*H* (Figure S2), guaranteeing the applicability of GN to select low-Δ*H* structure among numerous structural candidates.

With the same random seed, the first 200 steps for BO and RAS are the same in Figure 3(a). After that, BO show superior optimization performance than RAS. Most of BO-selected structures are located in low-Δ*H* configurations (exploitation) and some of them are in high-Δ*H* configurations (exploration). It is clearly seen that BO is able to find the ground state structures more efficient than RAS, which is also proven by the step-by-step structural animations shown in Supplementary Materials. The initial random structures do not make sense with Ca-Ca wrong bond and isolated unbonded S atoms as shown in Figure 3(b). Then the formation enthalpy quickly dropped and reasonable structures without wrong bonds are suggested as Figure 3(c). After 3000 steps, the four-coordination structure [Figure 3(d)] is found and after 5000 step, the ground-state RS structure [Figure 3(e)] is successfully predicted. Since the accuracy of current GN model is mildly inferior to state-of-the-art DFT calculations (Figure S2), the final structure in Figure 3(e) still have some distortions. Further BO steps may help to minimize this distortion (Figure S3). However, predicting a perfect RS structure remains a current challenge as the GN model is still inferior to accurately perceive little coordinate perturbation to total energies. Instead, we further used DFT calculations to relax the GN-BOSS predicted structures to obtain the precise crystal structures. To verify that the ground state crystal structures are derived from GN-BOSS rather than DFT relaxations. We adopted eighteen prototype structures of AB compounds that frequently appear in Inorganic Crystal Structure Database [34], as initial structures of CaS and only three of them were relaxed to RS structure (Figure S4). Those results further confirm that the ground state structure is derived from GN-BOSS approach but not DFT relaxation. This indicates that DFT relaxations help to find the precise location of a local minimum in structural space $(\{\mathbf{R}_i\},\mathbf{L})_{i=1,4}$ but cannot jump from one local minimum to another. Instead, GN-BOSS approach is able to find the basin around the global minimum and DFT calculation help to identify the exact location of minimum at that basin.

The GN-BOSS approach is then applied to structural searching for 23 other binary compounds. The corresponding $\Delta H$ evolution with BO/RAS steps have been shown in Figure S5-S7 and the results have been summarized in Table 1. It is seen that GN-BOSS can correctly and efficiently predict the experimental structures for all 24 compounds within 30 minutes by average via computational resource of only one CPU core. Notably, CsCl is commonly considered as an eight-coordination CsCl-type structure in experiments, while GN-BOSS predicts it as RS structure. In fact, the formation enthalpies of the two structures for CsCl are very close. In OQMD, the RS is more stable than CsCl-type structure by 47 meV per atom. In this sense, the GN-BOSS correctly reproduce the DFT results. This error is not by GN-BOSS approach but by the database. CsCl-type structure can be found when ten lowest-$\Delta H_f^{GN}$ structures within 7500 steps are selected, as shown in Table 2. Note that $\Delta H_f^{GN}$ of those ten structures for CsCl differ only by 32.5 meV/atom and there are at least three hundreds kinds of prototype structures for binary AB compounds [27]. In this sense, GN-BOSS is able to discern CsCl-type structure, demonstrating its strong ability to select the low-energy structures among large amount of structures candidates.

In conclusion, we combined graph network and Bayesian optimization as a GN-BOSS approach for crystal structure prediction without density functional theory calculations. GN-BOSS was then applied to successfully *predict* the crystal structures of typical I-VII (I=Li, Na, K, Rb, Cs; VII=F, Cl) and II-VI (II=Be, Mg, Ca, Sr, Ba, Zn, Cd; VI=O, S) compounds. The comparative studies show that GN-BOSS, although with mildly less accuracy to DFT results, is able to predict crystal structures from scratch without DFT. There are several directions for further improving current approach towards predicting more complicated and unknown structures in a more efficient way, such as reliable database and appropriate approaches for crystal structure characterization, structural searching and algorithm parallelization. This work may open a new avenue to data-driven crystal structural prediction without using the expensive DFT calculations during structural searching.

The work was supported by National Natural Science Foundation of China (grant No. 11974257), National Key Research and Development Program of China (grant No. 2020XXXXXXX), Jiangsu Distinguished Professorship, the Priority Academic Program Development of Jiangsu Higher Education Institutions (PAPD). DFT calculations were carried out at the National Supercomputer Center in Tianjin [TianHe-1(A)].

**Reference**

[1] L. Pauling, *THE PRINCIPLES DETERMINING THE STRUCTURE OF COMPLEX IONIC CRYSTALS*, J. Am. Chem. Soc. **51**, 1010 (1929).

[2] V. M. Goldschmidt, *Die Gesetze der Krystallochemie*, Naturwissenschaften **14**, 477 (1926).

[3] J. A. Van Vechten, *Quantum Dielectric Theory of Electronegativity in Covalent Systems. I. Electronic Dielectric Constant*, Phys. Rev. **182**, 891 (1969).

[4] J. C. PHILLIPS, *Ionicity of the Chemical Bond in Crystals*, Rev. Mod. Phys. **42**, 317 (1970).

[5] C. J. Pickard and R. J. Needs, *Ab Initiorandom Structure Searching*, J. Phys. Condens. Matter **23**, 053201 (2011).

[6] I. Sugden, C. S. Adjiman, and C. C. Pantelides, *Accurate and Efficient Representation of Intra-molecular Energy in Ab Initio Generation of Crystal Structures. I. Adaptive Local Approximate Models*, Acta Crystallogr. Sect. B Struct. Sci. Cryst. Eng. Mater. **72**, 6 (2016).

[7] J. Pannetier, J. Bassas-Alsina, J. Rodriguez-Carvajal, and V. Caignaert, *Prediction of Crystal Structures from Crystal Chemistry Rules by Simulated Annealing*, Nature **346**, 6282 (1990).

[8] D. M. Deaven and K. M. Ho, *Molecular Geometry Optimization with a Genetic Algorithm*, Phys. Rev. Lett. **75**, 288 (1995).

[9] C. W. Glass, A. R. Oganov, and N. Hansen, *USPEX—Evolutionary Crystal Structure Prediction*, Comput. Phys. Commun. **175**, 713 (2006).

[10] G. Trimarchi and A. Zunger, *Global Space-Group Optimization Problem: Finding the Stablest Crystal Structure without Constraints*, Phys. Rev. B **75**, 104113 (2007).

[11]     Y. Wang, J. Lv, L. Zhu, and Y. Ma, *Crystal Structure Prediction via Particle-Swarm Optimization*, Phys. Rev. B **82**, 094116 (2010).

[12]     Y.-Y. Zhang, W. Gao, S. Chen, H. Xiang, and X.-G. Gong, *Inverse Design of Materials by Multi-Objective Differential Evolution*, Comput. Mater. Sci. **98**, 51 (2015).

[13]     F. A. Faber, A. Lindmaa, O. A. von Lilienfeld, and R. Armiento, *Machine Learning Energies of 2 Million Elpasolite ( A B C 2 D 6 ) Crystals*, Phys. Rev. Lett. **117**, (2016).

[14]     C. J. Bartel, S. L. Millican, A. M. Deml, J. R. Rumptz, W. Tumas, A. W. Weimer, S. Lany, V. Stevanović, C. B. Musgrave, and A. M. Holder, *Physical Descriptor for the Gibbs Energy of Inorganic Crystalline Solids and Temperature-Dependent Materials Chemistry*, Nat. Commun. **9**, 1 (2018).

[15]     Z. Li, Q. Xu, Q. Sun, Z. Hou, and W.-J. Yin, *Thermodynamic Stability Landscape of Halide Double Perovskites via High-Throughput Computing and Machine Learning*, Adv. Funct. Mater. **29**, 1807280 (2019).

[16]     M. Tsubaki and T. Mizoguchi, *Quantum Deep Field: Data-Driven Wave Function, Electron Density Generation, and Atomization Energy Prediction and Extrapolation with Machine Learning*, Phys. Rev. Lett. **125**, 206401 (2020).

[17]     M. R. Carbone, M. Topsakal, D. Lu, and S. Yoo, *Machine-Learning X-Ray Absorption Spectra to Quantitative Accuracy*, Phys. Rev. Lett. **124**, 156401 (2020).

[18]     Y.-H. Liu and E. P. L. van Nieuwenburg, *Discriminative Cooperative Networks for Detecting Phase Transitions*, Phys. Rev. Lett. **120**, 176401 (2018).

[19]     A. Jain and T. Bligaard, *Atomic-Position Independent Descriptor for Machine Learning of Material Properties*, Phys. Rev. B **98**, (2018).

[20]     L. Ward, R. Liu, A. Krishna, V. I. Hegde, A. Agrawal, A. Choudhary, and C. Wolverton, *Including Crystal Structure Attributes in Machine Learning Models of Formation Energies via Voronoi Tessellations*, Phys. Rev. B **96**, (2017).

[21]     T. Xie and J. C. Grossman, *Crystal Graph Convolutional Neural Networks for an Accurate and Interpretable Prediction of Material Properties*, Phys. Rev. Lett. **120**, 145301 (2018).

[22]     C. Chen, W. Ye, Y. Zuo, C. Zheng, and S. P. Ong, *Graph Networks as a Universal Machine Learning Framework for Molecules and Crystals*, Chem. Mater. (2019).

[23]     C. J. Bartel, A. Trewartha, Q. Wang, A. Dunn, A. Jain, and G. Ceder, *A Critical Examination of Compound Stability Predictions from Machine-Learned Formation Energies*, Npj Comput. Mater. **6**, 97 (2020).

[24]     F. Aurenhammer, *Voronoi Diagrams - a Survey of a Fundamental Geometric Data*

*Structure*, ACM Comput. Surv. **23**, 345 (1991).

[25] C. H. Rycroft, *VORO++: A Three-Dimensional Voronoi Cell Library in C++*, Chaos Interdiscip. J. Nonlinear Sci. **19**, 041111 (2009).

[26] P. W. Battaglia, J. B. Hamrick, V. Bapst, A. Sanchez-Gonzalez, V. Zambaldi, M. Malinowski, A. Tacchetti, D. Raposo, A. Santoro, R. Faulkner, C. Gulcehre, F. Song, A. Ballard, J. Gilmer, G. Dahl, A. Vaswani, K. Allen, C. Nash, V. Langston, C. Dyer, N. Heess, D. Wierstra, P. Kohli, M. Botvinick, O. Vinyals, Y. Li, and R. Pascanu, *Relational Inductive Biases, Deep Learning, and Graph Networks*, ArXiv180601261 Cs Stat (2018).

[27] J. E. Saal, S. Kirklin, M. Aykol, B. Meredig, and C. Wolverton, *Materials Design and Discovery with High-Throughput Density Functional Theory: The Open Quantum Materials Database (OQMD)*, JOM **65**, 1501 (2013).

[28] A. Seko, H. Hayashi, K. Nakayama, A. Takahashi, and I. Tanaka, *Representation of Compounds for Machine-Learning Prediction of Physical Properties*, Phys. Rev. B **95**, (2017).

[29] C. W. Park and C. Wolverton, *Developing an Improved Crystal Graph Convolutional Neural Network Framework for Accelerated Materials Discovery*, Phys. Rev. Mater. **4**, 063801 (2020).

[30] J. Bergstra, D. Yamins, and D. Cox, *Making a Science of Model Search: Hyperparameter Optimization in Hundreds of Dimensions for Vision Architectures*, in *International Conference on Machine Learning* (PMLR, 2013), pp. 115–123.

[31] J. S. Bergstra, R. Bardenet, Y. Bengio, and B. Kégl, *Algorithms for Hyper-Parameter Optimization*, in *Advances in Neural Information Processing Systems 24*, edited by J. Shawe-Taylor, R. S. Zemel, P. L. Bartlett, F. Pereira, and K. Q. Weinberger (Curran Associates, Inc., 2011), pp. 2546–2554.

[32] T. Yamashita, N. Sato, H. Kino, T. Miyake, K. Tsuda, and T. Oguchi, *Crystal Structure Prediction Accelerated by Bayesian Optimization*, Phys. Rev. Mater. **2**, 013803 (2018).

[33] L. M. Ghiringhelli, J. Vybiral, S. V. Levchenko, C. Draxl, and M. Scheffler, *Big Data of Materials Science: Critical Role of the Descriptor*, Phys. Rev. Lett. **114**, (2015).

[34] G. Bergerhoff, R. Hundt, R. Sievers, and I. D. Brown, *The Inorganic Crystal Structure Data Base*, J. Chem. Inf. Model. **23**, 66 (1983).

**Table 1**. The experimental structures, GS structure in OQMD, GN-BOSS predicted ten low-energy structures within 7500 steps, GN-BOSS predicted GS structures and the spent BO step/ elapsed time for all twenty-four binary compounds. The analysis on 24 kinds of elapsed time lead to averaged time of 30 mins to predict one crystal structure (Figure S8).

| | Exp. Str. | GN-BOSS predicted ten lowest-energy structures in 7500 steps | GN-BOSS predicted most stable structure | Step | Elapsed time (min) |
|---|---|---|---|---|---|
| LiF | RS [$Fm\bar{3}m$] | 7 RS, 3 [$P6_3/mmc$] | RS | 3564 | 54 |
| NaF | RS [$Fm\bar{3}m$] | 10 [$P6_3/mmc$ | RS | 7985 | 99 |
| KF | RS [$Fm\bar{3}m$] | 8 RS, 2 [$R3m$] | RS | 17876 | 211 |
| RbF | RS [$Fm\bar{3}m$] | 2 RS, 8 [$P1$] | RS | 4034 | 22 |
| CsF | RS [$Fm\bar{3}m$] | 2 RS, 2 [$P6_3/mmc$], 2 [$P\bar{6}m2$], 1 [$R3m$], 2 [$P6_3/mmc$], 1 [$P1$] | RS | 3746 | 17 |
| LiCl | RS [$Fm\bar{3}m$] | 8 RS, 2 ZB, | RS | 17231 | 284 |
| NaCl | RS [$Fm\bar{3}m$] | 2 RS, 4 [$R3m$], 2 [$P6_3/mmc$] | RS | 6042 | 61 |
| KCl | RS [$Fm\bar{3}m$] | 8 RS, 2 [$P6_3/mmc$] | RS | 3639 | 23 |
| RbCl | RS [$Fm\bar{3}m$] | 9 RS, 1 [$P1$] | RS | 5930 | 41 |
| CsCl | RS [$Fm\bar{3}m$] | 3 RS, 2 CsCl-type, 5 [$R3m$] | RS | 12246 | 95 |
| BeO | WZ [$P6_3mc$] | 1 WZ, 7 ZB, 2 [$R3m$] | WZ/ZB | 3656 | 96 |
| MgO | RS [$Fm\bar{3}m$] | 5 RS, 1 [$P6_3/mmc$], 1 [$P6_3/mmc$], 3 [$Cmc2_1$] | RS | 4118 | 55 |
| CaO | RS [$Fm\bar{3}m$] | 10 RS | RS | 1881 | 12 |
| SrO | RS [$Fm\bar{3}m$] | 3 RS, 6 [$P6_3/mmc$], 1 [$Cmc2_1$] | RS | 7745 | 74 |
| BaO | RS [$Fm\bar{3}m$] | 5 RS, 2 WZ, 1 [$Cm$], 1 [$P6_3/mmc$], 1 [$Cmc2_1$] | RS | 7450 | 60 |
| ZnO | WZ [$P6_3mc$] | 10 WZ | WZ | 8895 | 161 |
| CdO | RS [$Fm\bar{3}m$] | 1 RS, 1 ZB, 7 [$P1$], 1 [$P6_3/mmc$] | RS | 6398 | 93 |
| BeS | ZB [$F\bar{4}3m$] | 1 [$C2/m$], 1 [$P6_3/mmc$], 7 [$P\bar{1}$], 1 [$Cm$] | ZB | 594 | 5 |
| MgS | RS [$Fm\bar{3}m$] | 8 RS, 1 ZB, 1 [$P6_3/mmc$] | RS | 5198 | 50 |
| CaS | RS [$Fm\bar{3}m$] | 10 RS | RS | 4047 | 31 |
| SrS | RS [$Fm\bar{3}m$] | 7 RS, 3 [$P\bar{6}m2$] | RS | 6944 | 54 |
| BaS | RS [$Fm\bar{3}m$] | 10 RS | RS | 4631 | 30 |
| ZnS | ZB [$F\bar{4}3m$] | 9 ZB, 1 WZ | ZB/WZ | 3281 | 25 |
| CdS | WZ [$P6_3mc$] | 2 WZ, 2 ZB, 1 RS, 1 [$PMMN$], 4 [$P1$] | WZ | 5290 | 38 |

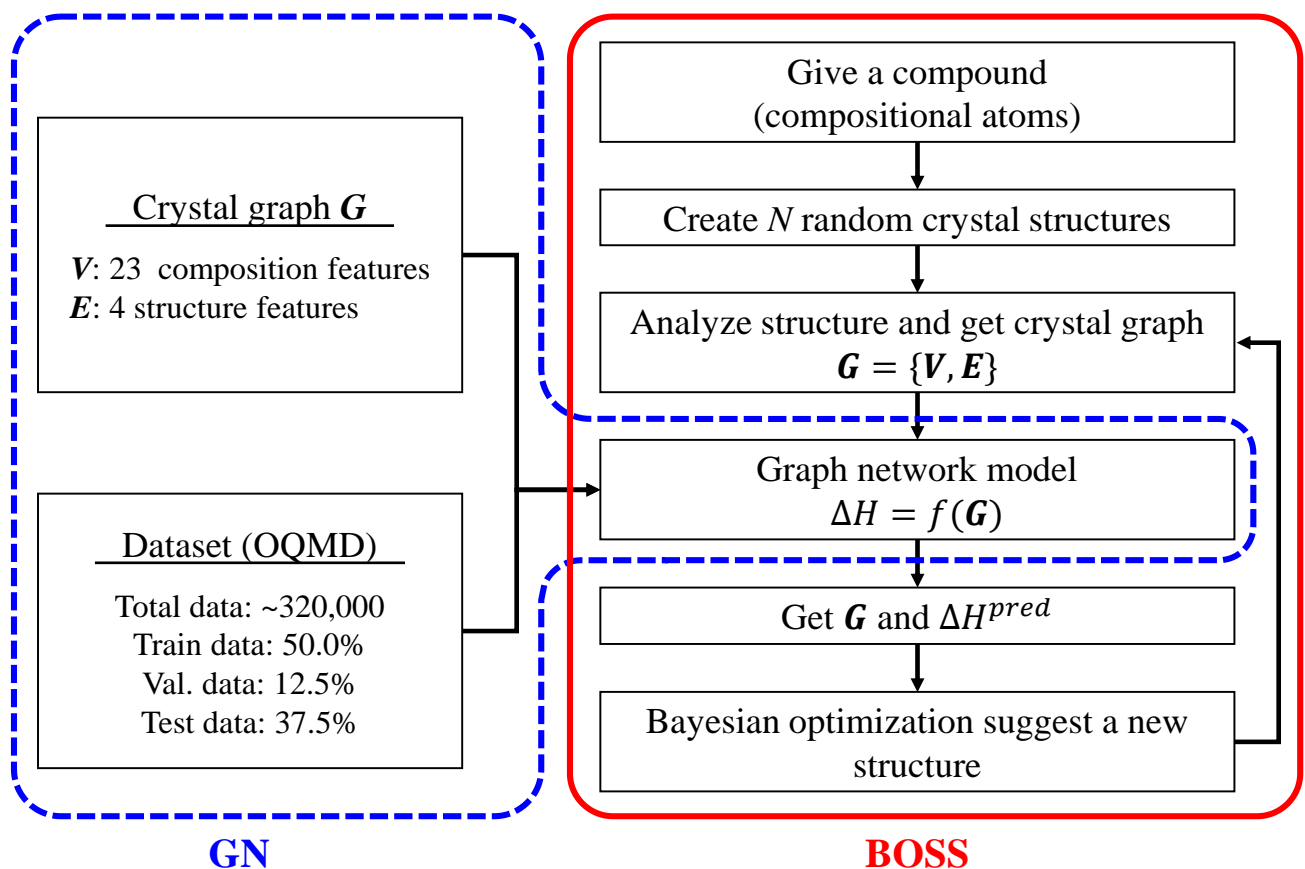

**Figure 1.** The flow-chart of GN-BOSS approach.

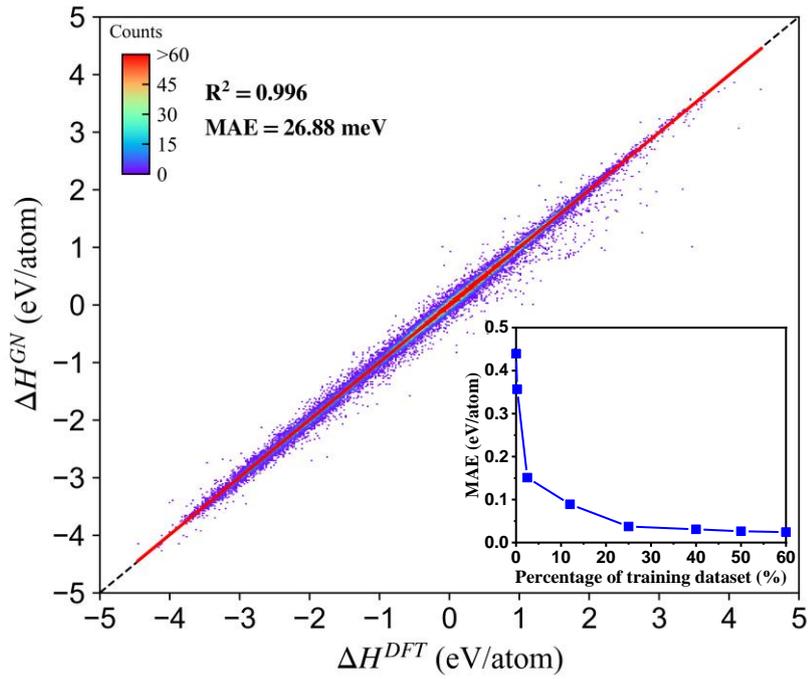

**Figure 2.** GN-predicted and DFT-calculated formation enthalpies for ~320000 crystals in OQMD. The mean absolute error as a function of percentage of training crystals.

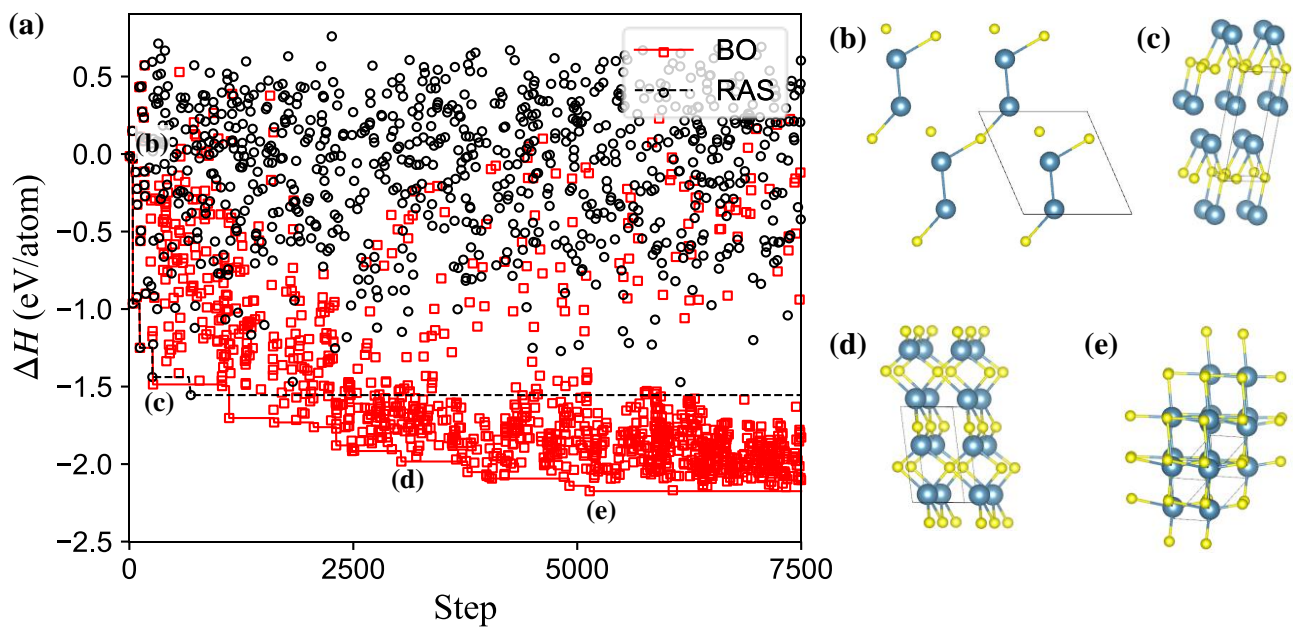

**Figure 3.** The process of GN-BOSS approach, in comparison to RAS, to search the crystal structure of CaS. Typical structures derived from GN-BOSS process are shown in (b-e).